\documentclass [aip,jcp,jmp, amsmath,amssymb]{revtex4-1}
\usepackage{graphicx}
\usepackage[version=3]{mhchem}
\usepackage{bm}

\begin{document}

\title[]{Entropic estimate of cooperative binding of substrate on a 
single oligomeric enzyme: An index of cooperativity}

\author{ Kinshuk Banerjee, Biswajit Das and Gautam Gangopadhyay}

\email{gautam@bose.res.in}

\affiliation{S.N.Bose National Centre For Basic Sciences\\Block-JD, 
Sector-III, Salt Lake, Kolkata-700098, India.}

%\date{\today}

\begin{abstract}

Here we have systematically studied the cooperative binding of substrate 
molecules on the active sites of a single oligomeric enzyme in a chemiostatic condition. 
The average number of 
bound substrate and the net velocity of the enzyme catalyzed reaction 
are studied by the formulation of stochastic 
master equation for the cooperative binding classified here as 
spatial and temporal. 
We have estimated the entropy production for the cooperative binding 
schemes 
based on single trajectory analysis using a kinetic Monte Carlo 
technique. It is found that the total as well as the medium entropy 
production show the same generic diagnostic signature for detecting the 
cooperativity, usually characterized in terms of the net velocity of 
the reaction. This feature is also found to be valid for the 
total entropy production rate at the nonequilibrium steady state. 
We have introduced an index of cooperativity, C, defined in terms of the 
ratio of the surprisals or equivalently, the stochastic system entropy 
associated with 
the fully bound state of the cooperative and non-cooperative cases. 
The criteria of cooperativity in terms of C is compared with 
 that of the Hill coefficient and 
gives a microscopic insight on the cooperative binding of substrate on a 
single oligomeric enzyme which is usually characterized by 
macroscopic reaction rate.

\end{abstract}

\keywords{Oligomeric enzyme kinetics, Cooperativity, Stochastic thermodynamics, Single trajectory}

\maketitle

{\section{Introduction}}

Conventional thermodynamics
at or near equilibrium needs serious modification to accommodate the events 
of single molecular processes 
as well as nano-systems which are generally in states far 
away from equilibrium \cite{collin,jarzynski1,liphardt,wang,carberry,
ritort1,ritort2,ritort3,ritort4,ritort5}. 
The single molecule study is very important in biological systems because most 
of the processes in cell 
are taking place on the level of a single or a few molecules. 
The non-equilibrium feature is mainly developed within a cell due to the  
mechanical or chemical stimuli which runs the metabolism through the driven  
chemical reactions\cite{seifert1,seifert2}.
Quantitative measure of fluctuations 
\cite{evans1,cohen1,spohn,crooks1,sasa,gallavotti,jarzynski2,sevick} in 
small systems are possible over short periods of time that 
allow the thermodynamic concepts to be applied to such finite systems. 
A crucial concept in the statistical description of a nonequilibrium small system is 
that of a single trajectory or path\cite{ritort4,crooks1,jarzynski2,evans1} 
and defining the entropy of the system for 
such a single trajectory 
allows one to formulate the second law of thermodynamics at the stochastic level
\cite{seifert4,seifert5,seifert6,seifert7}. 
The trajectory-based entropy 
production\cite{seifert4,seifert5,seifert6} has been 
successfully applied to various systems, for example, 
single bio-molecular reactions\cite{seifert1}, 
chemical reaction networks\cite{seifert5} and 
driven colloidal particles\cite{seifert6}.

Enzyme kinetics is a very important process in cellular metabolism where 
nonequilibrium feature is developed due to the imbalanced chemical 
reactions and the presence of chemiostatic condition prevents the 
reaction system to attain equilibrium\cite{qian1,qian2,qian3,bdas}. In a chemiostatic 
condition, substrate and product are maintained at constant concentrations
 by continuous influx of the substrate  and withdrawing the product
from the system. Under this condition, the reaction system 
reaches a nonequilibrium steady state (NESS) \cite{qian1,qian2,qian3} which
is characterized by a non-zero total entropy production rate. 
Single molecule enzyme kinetics \cite{xie1,xie2} is theoretically studied  
using the stochastic master equation approach\cite{ge1,qian4} as well
as by the stochastic single trajectory analysis \cite{seifert1,seifert2}. 
Now most of the enzymes found in enzymology are oligomeric in nature 
consisting of two or more subunits usually 
linked to each other by non-covalent interactions\cite{palmer}. 
Possibility of interaction between the subunits during the 
substrate binding process can give rise to different cooperative phenomena
\cite{palmer,ricard,hammes,goldbeter1}. 
Positive cooperativity is said to occur when the binding of one 
substrate molecule with a subunit increases the affinity 
of further attachment of the substrate  to another 
subunit\cite{palmer,ricard,weber}. 
In the case of negative cooperativity, attachment of a substrate molecule
to one subunit 
decreases the tendency of further attachment of the substrate molecules 
to other subunits\cite{levitzki,abeliovich}. 
These types of cooperativity based on the affinity of the substrate 
binding belong to the class of allosteric cooperativity
\cite{weber,ricard,hammes,goldbeter1}. 
There is another type of cooperativity, termed as 
temporal cooperativity\cite{qian5}, reflected in the zero-order 
ultra sensitivity of the phosphorylation-dephosphorylation cycle 
which is shown to be mathematically equivalent to the allosteric cooperativity 
\cite{ge1}. 
Beside allosterism, cooperativity has been studied in monomeric enzymes 
with only a single substrate binding site. This has led to the important 
concept of hysteretic \cite{ricard,qian4,cardenas} and mnemonic
 enzymes\cite{frieden, neet}. 
These two types of enzymes show the cooperativity phenomena due to the 
slow conformational disorder of the active site \cite{qian4}.

In this paper, we have 
studied the entropy production in the kinetics of a single 
oligomeric enzyme which shows cooperativity with respect to the 
substrate binding. 
Here we have classified the cooperativity phenomena according to the nature of 
the different substrate binding mechanisms, namely, 
sequential and independent, as detailed by Weiss\cite{weiss}. 
%spatial and non-spatial or temporal. 
In sequential binding, 
the adjacent sites of the oligomeric enzyme are 
successively occupied by the substrate molecules. 
So the substrate-bound states of the 
system are actually adjacent in space and 
hence we denote the cooperativity arising out of this binding protocol 
as the spatial cooperativity.  For the sequential mechanism, the 
first binding site, {\it i.e}, the first subunit of the oligomeric enzyme 
must be filled in order for the second site to become 
occupied by the substrate, 
as if the substrate molecules have been stacked 
on top of each other at their binding sites\cite{weiss}. 
This type of binding can be 
relevant to an ion transporter, such as the Na-K 
pump\cite{gadsby}.
The other class is called temporal cooperativity 
which can occur due to the independent binding 
of the substrate molecules to any one of the subunits at a particular
time without any specific spatial arrangement.
Here the substrate-bound sites are not physically neighboring 
in the enzyme \cite{qian5} but the global state of the system is defined
 in terms of the total occupancy of the  overall sites at a 
particular instant of time. This type of binding can be observed in 
multimeric proteins with individual binding sites 
located on different subunits, such as ligand gated ion channels 
or ligand gated enzymes\cite{weiss}.
Here we have theoretically studied 
the cooperative behavior solely from the viewpoint of the substrate binding 
mechanism and not in terms of the active and inactive enzyme conformations or 
the actual structural details of the enzyme that can lead to such mechanisms 
\cite{monod,koshland}. 
To study the bulk kinetics of allosteric enzymes Monod, 
Wyman and Changeux (MWC) in 1965 and Koshland, Nemethy and Filmer (KNF) in 
1966 put forward models to account for cooperative binding. 
Generally $\rm `Sequential$' is used  as a term  for a classical distinction
between multi-step binding models, as for instance
differentiating between the KNF and MWC models in terms of the variation of
the substrate binding rates in each successive step. 
So the term ‘sequential’ in KNF model  
should not be confused with the term ‘sequential’ used in our approach
\cite{weiss,bindslev}. We have constructed the 
master equations for each class of substrate binding. 
Time evolution of such cooperative systems can be described by suitably 
applying a kinetic Monte Carlo technique\cite{gillespie1,gillespie2}. 
Here we have applied this algorithm to calculate the 
total, medium and system entropy production along a single trajectory 
for such cooperative systems as a function of the substrate concentration  over a time interval where finally the 
system reaches a nonequilibrium steady state (NESS) 
and 
then determined the ensemble average quantities over many realizations 
of such trajectories. 
We show the correspondence between the evolution of the total 
and the medium entropy production with the 
average substrate binding and net 
velocity of the reaction in the context of detection of the 
cooperative behavior. Similarly this correspondence is also studied for the 
total entropy production rate at the NESS. 
The system entropy production is thoroughly studied 
in terms of the substrate binding probabilities 
for the different classes 
of cooperative systems considered. We have introduced a 
cooperativity index, C defined  
in terms of the stochastic system entropy 
to understand the nature of the cooperativity.

Layout of the paper is as follows. In Section II, we have given the 
master equations and their steady state solutions to describe the spatial 
and temporal cooperative binding mechanisms and the corresponding entropy
production rates.  
In Section III, numerical results of entropy production and 
cooperative kinetics is discussed. In Section IV, we have 
discussed on measures of cooperativity and introduced an index of 
cooperativity. 
Then the paper is concluded in the Section V.

\section{Cooperative binding, master equation and entropy 
production rate}

In this section, we have first classified the cooperativity of a single 
oligomeric enzyme on the basis of the  nature of the enzyme-substrate 
binding and then proposed a stochastic description for each class in 
terms of a one-dimensional random walk problem. Here we have provided 
a master equation approach for the description of spatial and temporal 
cooperativity which is suitable for the calculation of entropy production.

\subsection{ Classes of cooperativity: spatial and temporal substrate  
binding}

Here we have considered that the substrate molecules can bind to the subunits 
of the oligomeric enzyme sequentially or independently 
as already discussed in the Introduction section. 
 In the oligomeric enzyme kinetics reaction,
 the substrate molecules bind to the subunits of the oligomeric enzyme 
 in a stepwise manner with different affinity which was first proposed
 by Adair to explain the cooperativity phenomenon observed in the 
oxygen-binding to the hemoglobin at equilibrium\cite{adair}. 
If an oligomeric enzyme consists of  $\rm n_{T}$ number of homo or hetero
 type of subunits, then at the chemiostatic condition the substrate-binding 
scheme of the enzyme can be written as, 
\begin{equation}
\ce{E  <=>[\ce{K_{1}}^{(0)}][\ce{K_{2}}^{(1)}]  ${\ce{ES_{1}}}$ <=>[\ce{K_{1}}^{(1)}][\ce{K_{2}}^{(2)}] ${\ce{ES_{2}}}$  <=>[\ce{K_{1}}^{(2)}][\ce{K_{2}}^{(3)}] 
${......}$ <=>[\ce{K_{1}}^{(n-1)}][\ce{K_{2}}^{(n)}] ${\ce{ES_{n}}}$ <=>[\ce{K_{1}}^{(n)}][\ce{K_{2}}^{(n+1)}] ${......}$ <=>[\ce{K_{1}}^{(n_{T}-2)}][\ce{K_{2}}^{(n_{T}-1)}] ${\ce{ES_{n_{T}-1}}}$  <=>[\ce{K_{1}}^{(n_{T}-1)}][\ce{K_{2}}^{(n_{T})}] ${\ce{ES_{n_{T}}}}$}.
\label{e2esn}
\end{equation}
Here $\rm ES_{n}$ represents  the conformational state of the oligomeric enzyme
 in which n number of subunits are occupied by the substrate molecules.
$\rm {{ K}_{1}}^{(n-1)}$ and $\rm {{K}_{2}}^{(n)}$
are designated as the total formation and total dissociation rate constants
 in the n-th reaction step, respectively.

The above scheme of substrate binding of an oligomeric enzyme can be viewed
as a generalization of the kinetics of an enzyme having a single subunit
given by 
$$ \ce{E +S <=>[\ce{k_1}'][\ce{k_{-1}}] ${\ce{ES}}$ <=>[\ce{k_{-2}}][\ce{k_{2}}'] ${\ce{E + P}}$},$$ 
which can be further simplified as 
\begin{equation}
\ce{E  <=>[\ce{K_{1}}][\ce{K_{2}}]  ${\ce{ES}}$}.
\label{e2es}
\end{equation}
Here $\rm { K}_{1}=(k_{1}+k_{2})$ and $\rm {K}_{2}=(k_{-1}+k_{-2})$,
are designated as the total formation and total dissociation rate
constants of ES, respectively.
The pseudo first-order rate constants  are written as
 $\rm {k}_{1}={k}_1^{'}[ S]$ 
and $\rm {k}_{2}={k}^{'}_{2}[P]$ where $\rm [S]$ and $\rm [P]$ are
the constant substrate and product concentration in the chemiostatic condition. 
Hence the site-dependent total formation and dissociation rate constants in the 
case of the oligomeric enzyme kinetics are similarly defined as
 $\rm {{K}_{1}}^{(n-1)}=({k_{1}}^{(n-1)}+{k_{2}}^{(n-1)})$ and 
$\rm {{K}_{2}}^{(n)}=({k_{-1}}^{(n)}+{k_{-2}}^{(n)})$ where
$\rm {k_{1}}^{(n-1)}= {k_{1}'}^{(n-1)}[S]$ and 
$\rm {k_{2}}^{(n-1)}= {k_{2}'}^{(n-1)}[P]$.

\begin{figure}
\centering
\rotatebox{270}{
\includegraphics[width=10cm,keepaspectratio]{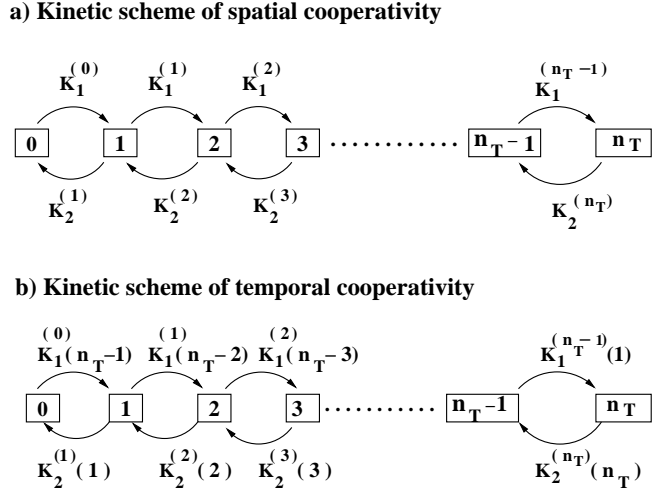}}
\caption{Kinetic schemes for (a) spatial and (b) temporal cooperativity with 
site-dependent binding and dissociation rate constants. The numbers 
in the square boxes denote the number of occupied sites.
In spatial
cooperativity (a), the forward and backward transition probabilities 
depend only on 
the total formation and dissociation rate constants, respectively. 
For temporal
cooperativity (b), the forward transition probability depends on 
the total formation rate constant and the number of 
unoccupied sites, whereas the backward transition probability depends on the 
total dissociation rate constant and the number of occupied sites.}
\label{fig_1}
\end{figure}

The dynamics of the substrate binding mechanisms are quantified by counting
 the number of occupied sites present in the oligomeric enzyme at a particular 
instant of time. If at time t, 
`$\rm n$' number of occupied sites are present in the oligomeric enzyme 
(the state $\rm ES_{n}$)
then at time $\rm t+dt$, the number of occupied sites may be increased or 
decreased by one unit due to the occurrence of 
a formation or a dissociation reaction. 
During the time evolution, the number of occupied sites 
is a fluctuating quantity. Therefore, the system performs a one-dimensional 
random walk along the finite number of states where  state-$\rm n$ of the
 system is equivalent to the conformational state $\rm ES_{n}$, 
 as shown in figure \ref{fig_1}. 
%The indices in the square boxes in figure 1 represent the states of the system. 

The kinetic scheme of the spatially cooperative system is given in 
figure \ref{fig_1}(a). As the subunits get occupied 
sequentially starting from subunit-1, so when we say that 
the $\rm n$-th subunit is occupied it automatically means that 
$\rm n$ number of sites are occupied in total. 
Here the forward and the backward transition probabilities 
 depend only on  
the total formation and dissociation rate constants $\rm K_{1}^{(n)}$ 
and $\rm K_{2}^{(n)}$, respectively which are generally site-dependent. 
This is so because after the filling of one subunit, there is no other 
choice for the next substrate molecule but to fill up the next adjacent 
subunit and as this is true for all the subunits, there is no combinatorial 
term in the transition probability. 
 
The kinetic scheme for the temporal cooperativity is shown 
in figure \ref{fig_1}(b). 
Here the substrate molecules can bind independently with any one of the 
$\rm n_{T}$ number of subunits. The state-$\rm n$ of the system represents 
$\rm n$-number of occupied sites of the enzyme. 
In this mechanism, the forward transition probability of 
the n-th state at time t is given by the product of  
the total formation rate constant $\rm K_{1}^{(n)}$ with the number of 
distinct combinations of unoccupied sites present at that time. 
Similarly, the backward transition probability 
of the same state is the product of the total 
dissociation rate constant, $\rm K_{2}^{(n)}$ 
and the number of distinct combinations
of occupied sites present at time t (see figure \ref{fig_1}(b)). 
Here these rate constants  are taken to be site-dependent in general. 
If all the rate constants are site-independent, 
then the system will be non-cooperative. 
The main difference of the sequential binding scheme from the 
independent one is as follows: {\it for the sequential 
scheme, the system will show spatial cooperativity in substrate
 binding even when the formation and dissociation rate constants
 are not site-dependent.}

\subsection{Master equations }
    
For the time-dependent description of the spatial cooperativity, 
we have constructed the 
corresponding master equation for this cooperativity mechanism as 
\begin{equation}
\frac{\partial \rm P_{sp}(n,t)}{\partial t}= \rm K_{1}^{(n-1)}P_{sp}(n-1;t)+ 
 \rm K_{2}^{(n+1)}P_{sp}(n+1;t)-\rm (K_{1}^{(n)}+K_{2}^{(n)})P_{sp}(n;t),
\label{meqseq}
\end{equation}
with $\rm K_{1}^{(-1)}=\rm K_{2}^{(0)}=\rm K_{1}^{(n_{T})}=\rm K_{2}^{(n_{T}+1)}
= 0$ to match the boundary terms. 
Here, $\rm P_{sp}(n,t)$ is the probability of having $n$ number of occupied 
sites at time t. We have given an analytical expression for the 
 solution of the master equation by setting 
$\rm \frac{\partial P_{sp}(n,t)}{\partial t} = 0 $.
The steady state distribution of the spatial cooperativity is given by   
\begin{equation}
\rm  P_{sp}^{\rm ss}(n)= \frac{\prod_{j=0}^{n-1}X^{(j)}}{\sum_{n=0}^{n_{T}}
\prod_{j=0}^{n-1}X^{(j)}},
\label{stdistrispace}         
\end{equation}
where $\rm X^{(j)}=\frac{\rm K_{1}^{(j)}}{\rm K_{2}^{(j+1)}}=
\frac{k_{1}^{'(j)}[S]+k_{2}^{(j)}}{k_{-1}^{(j+1)}+k_{-2}^{(j+1)}}$ with 
$\rm  j=0,1,...,(n_{\rm T}-1)$. 
Here the steady state is actually a nonequilibrium steady state (NESS) as 
already discussed. If $\rm X^{(j)}=X \forall j$, the NESS probability 
distribution becomes 
\begin{equation}
\rm  P_{sp}^{\rm ss}(n) = \frac{X^{n}(1-X)}{1-X^{(n_{\rm T}+1)}},
\label{geometricdistribution}    
\end{equation}
which is a geometric distribution. 
The average population of the occupied sites at the NESS for 
$\rm X^{(j)}=X \forall j$ 
is given by 
 \begin{equation}
\rm \left\langle n\right\rangle =\sum_{n=0}^{n_{T}} n P_{sp}^{\rm ss}(n)
 = \frac{X(1-(n_{T}+1)X^{n_{T}}+n_{T}X^{n_{T}+1})}{(1-X)(1-X^{n_{T}+1})}.
 \label{avgpopspace}
 \end{equation}
   
For temporal cooperativity, the master equation is written as
$$\rm   \frac{\partial \rm P_{temp}(n,t)}{\partial t}= K_{1}^{(n-1)}(n_{T}-n+1)
P_{temp}(n-1;t)+  K_{2}^{(n+1)} 
(n+1)P_{temp}(n+1;t)$$
\begin{equation}
\rm -K_{1}^{(n)}(n_{T}-n)P_{temp}(n;t)- K_{2}^{(n)}n P_{temp}(n;t),
\label{meqindp}
\end{equation}
again with $\rm K_{1}^{(-1)}= K_{2}^{(0)}= K_{1}^{(n_{T})}= K_{2}^{(n_{T}+1)}= 0 $. 
Solving this master equation at the NESS, we can obtain the probability 
distribution as    
\begin{equation}
 \rm P_{temp}^{\rm ss}(n)= \frac{\left( \begin{array}{ccc} n_{T} \\ n\end{array} \right)\prod_{j=0}^{n-1} X^{(j)}}{\sum_{n=0}^{n_{T}}\left( \begin{array}{ccc} n_{T} 
\\ n\end{array} \right)\prod_{j=0}^{n-1} X^{(j)}},
    \label{stdisttemp}
   \end{equation} 
where $\rm X^{(j)}=\frac{K_{1}^{(j)}}{K_{2}^{(j+1)}}$ as already defined 
with $\rm j=0,1,...,(n_{T}-1).$  
The average number of occupied sites at the NESS is simply expressed as
\begin{equation}
\rm \left\langle n \right\rangle =\frac{\sum_{n} n\left
( \begin{array}{ccc} n_{T} \\ 
n\end{array} \right)\prod_{j=0}^{n-1} X^{(j)}}{\sum_{n=0}^{n_{T}}\left
( \begin{array}{ccc} n_{T} \\ 
n\end{array} \right)\prod_{j=0}^{n-1} X^{(j)}}.
\label{navrgind}
\end{equation}

Now positive cooperativity in this 
scenario means a higher affinity of a second substrate molecule to 
attach with the oligomeric enzyme compared to that of the first substrate 
molecule which is already bound and so on. Therefore, in this case, 
the successive binding affinity of the substrate molecule increases.
So naturally here we take the binding rate 
constants, $\rm k_1^{(n)}$  as follows \cite{weiss,tan}
  \begin{equation}
\rm  k_{1}^{(0)} < k_{1}^{(1)}....< k_{1}^{(n)} < k_{1}^{(n+1)} <....< k_{1}^{(n_{T}-1)}.
 \label{poscop}
  \end{equation}
Here the site-dependent overall association rate constant $\rm K_{1}^{(n)}$ is 
defined as 
$\rm K_{1}^{(n)}= k_{1}^{(n)}+ k_{2}^{(n)}$ and similarly, the overall 
site-dependent dissociation rate constant is 
written as $ \rm K_{2}^{(n)}= k_{-1}^{(n)}+ k_{-2}^{(n)}$. 
We take the rate constants $\rm k_{-1}^{(n)} $, $\rm k_{2}^{(n)}$ and 
$\rm k_{-2}^{(n)}$ to be site-independent.
This is due to the fact that to get a cooperative behavior 
for the independent binding case, 
it is not necessary to take all the rate constants of the 
reaction system to be site-dependent that will also 
make the results obtained hard to analyze. 
Then the site-dependent quantities $\rm X^{(j)}$ for positive 
cooperativity maintain the relation:
 \begin{equation}
%\rm  K_{1}^{(0)} < K_{1}^{(1)}....< K_{1}^{(n)} < K_{1}^{(n+1)} <....< K_{1}^{(n_{T}-1)}.
\rm  X^{(0)} < X^{(1)}....< X^{(n)} < X^{(n+1)} <....< X^{(n_{T}-1)}.
\label{asopco}
 \end{equation}

Similarly, negative cooperativity arises 
as a second substrate molecule binds to 
the oligomeric enzyme with a lower affinity than that of the first 
substrate molecule. Therefore,  the substrate 
binding reaction rate constants for different sites obey the 
inequalities    
  \begin{equation}
\rm   k_{1}^{(0)} > k_{1}^{(1)}....> k_{1}^{(n)} > k_{1}^{(n+1)} >...> k_{1}^{(n_{T}-1)}.
\label{neco}
\end{equation}
Then taking the rate constants $\rm k_{-1}^{(n)} $, $\rm k_{2}^{(n)}$ 
and $\rm k_{-2}^{(n)}$ as site-independent constants, we have 
%the overall association rate constants obey the relation:
 \begin{equation}
% \rm  K_{1}^{(0)} > K_{1}^{(1)}....> K_{1}^{(n)} > K_{1}^{(n+1)} >....> K_{1}^{(n_{T}-1)}.
\rm  X^{(0)} > X^{(1)}....> X^{(n)} > X^{(n+1)} >....> X^{(n_{T}-1)}.
 \label{asoneco}
 \end{equation}

If all the association and dissociation rate constants are site-independent, 
then the enzyme becomes non-cooperative. Thus the steady state 
distribution, Eq.(\ref{stdisttemp}) reduces to 
  \begin{equation}
\rm  P^{ss}(n)= \left(\begin{array}{ccc} n_{T} \\ n\end{array} \right)
 \frac{X^{n}}{(1+X)^{n_{T}}},
\label{binom}
 \end{equation} 
where $\rm X= \frac{K_{1}}{K_{2}}$. By inserting the value of X, the above 
equation can be written as a binomial distribution given by
\begin{equation}
\rm  P^{ss}(n)= \left( \begin{array}{ccc} n_{T} \\ n\end{array} \right)
\left( \frac{K_{1}}{K_{1}+K_{2}}\right) ^{n}
 \left( \frac{K_{2}}{K_{1}+K_{2}}\right) ^{(n_{T}-n)}= P^{(bino)}(n).
\label{binom1}
\end{equation} 
This is expected, as in the absence of any cooperativity, the distribution of 
the occupied sites must follow a binomial distribution.
So for a system with  no cooperativity, 
the average number of occupied sites at the NESS is 
\begin{equation}
\rm  \left\langle n \right\rangle = n_{T}\left( \frac{X}{1+X}\right) 
= n_{T}\left( \frac{K_{1}}{K_{1}+ K_{2}}\right)
\label{navrgno}
\end{equation} 
and the average number of unoccupied sites is 
\begin{equation}
\rm  \left\langle n_{\rm T}-n \right\rangle =n_{T}\left( \frac{K_{2}}{K_{1}+ K_{2}}\right). 
\end{equation}
We mention that, in addition to the overall association and dissociation rate 
constants being site-independent, if 
the rate constant $\rm k_{2}$ is also negligibly small, then the enzyme 
kinetics becomes simply the Michaelis-Menten type. 
If  $\rm k_{2}^{(j)}$ ($\rm j=0,\ldots,(n_T-1)$) is taken
to be much less than the other rate constants, then we have 
\begin{equation}
\rm X^{(j)}=\frac{[S]}{K_M^{(j)}},
\label{Kmstpwise}
\end{equation}
where $\rm K_M^{(j)}=\frac{k_{-1}^{(j+1)}+k_{-2}^{(j+1)}}{k_{1}^{'(j)}}$ 
can be described as the stepwise Michaelis-Menten constant.

\subsection{ Entropy production rates}

The system entropy is defined in terms of the Shannon entropy as 
\begin{equation}
\rm S_{\rm sys}(t)=-k_{B}\sum_{n}P(n,t){\rm ln} P(n,t),
\label{shannon}
\end{equation}
where $\rm P(n,t)$ is the probability of having n number of occupied states at 
time t with $\rm P(n,t)\equiv P_{sp}(n,t)$ or 
$\rm P(n,t)\equiv P_{temp}(n,t).$ 
Here we set the Boltzmann constant, $\rm k_{B}=1$. 
Using the master equation, one can get the system entropy production rate 
\cite{nicolis1,nicolis2,gaspard1,gaspard2} 
as 
$$\rm \dot{S}_{sys}(t)=\frac{1}{2}\sum_{n,\mu}[w_{\mu}(n-\nu_{\mu}|n)
P(n-\nu_{\mu},t)
-w_{-\mu}(n|n-\nu_{\mu})P(n,t)]$$
\begin{equation}
\rm \times{\rm ln}\frac{P(n-\nu_{\mu},t)}{P(n,t)}.
\label{eratesys}
\end{equation} 
Here the state of the system can change by any one of the four reactions,  
denoted with index $\mu$, 
via which the substrate and the product molecules can bind with the 
enzyme sites and detach. 
They are given as: 
(1) $\rm ({ES_{n} +S}) \stackrel{{k_1^{(n)}}}{\rightarrow} (ES_{n+1})$ $(\mu=1)$, 
(2)$\rm (ES_{n}) \stackrel{k_{-1}^{(n)}}{\rightarrow} {(ES_{n-1}+S)}$ $(\mu=-1)$, 
(3)$\rm (ES_{n}) \stackrel{{k_{-2}^{(n)}}}{\rightarrow} {(ES_{n-1} + P)}$ $(\mu=-2)$ and 
(4)$\rm {(ES_{n} +P)} \stackrel{{k_{2}^{(n)}}}{\rightarrow} (ES_{n+1})$ $(\mu=2)$. 
Here $\rm \nu_{\mu}$ is designated 
as the stoichiometric coefficient of the $\rm \mu$-th reaction with 
rate constant $\rm k_{\mu}$ where $\rm \nu_{\mu}=1$ with $\rm \mu > 0 $ and $\rm  -\nu_{\mu}=1$ 
with $\rm \mu<0$.  
The transition probabilities are defined as follows 
%when $\mu$ is positive
$$\rm w_{\mu}(n-\nu_{\mu}|n)=k_{\mu}^{(n-\nu_{\mu})}(n_{T}-(n-\nu_{\mu})), \mu > 0$$
and 
%when $\mu$ is negative,
\begin{equation}
\rm w_{\mu}(n-\nu_{\mu}|n)=k_{\mu}^{(n-\nu_{\mu})}(n-\nu_{\mu}), \mu<0.
\label{propen}
\end{equation}

We have assumed ideal reservoir(surroundings) with no 
inherent entropy production except through the boundaries of the system. 
The system entropy production rate(epr) can be split as\cite{nicolis1} 
%%%%%%%%%%%%%%%%%%%%%%%%%%%%%%%
\begin{equation}
\rm \dot{S}_{sys}(t)=\dot{S}_{tot}(t)-\dot{S}_{m}(t).
\label{2ndlaw}
\end{equation}
Here the first term in the r.h.s. of equation({\ref{2ndlaw}}) 
gives the total entropy production rate and the second term denotes 
the medium entropy production rate due to the 
entropy flux into the surroundings. 
Therefore the total and medium entropy production rates
are defined as 
$$\rm \dot{S}_{tot}(t)=\frac{1}{2}\sum_{n,\mu}[w_{\mu}(n-\nu_{\mu}|n)
P(n-\nu_{\mu},t)-w_{-\mu}(n|n-\nu_{\mu})P(n,t)]$$
\begin{equation}
\rm \times\hspace{0.1cm}{\rm ln}\frac{w_{\mu}(n-\nu_{\mu}|n)P(n-\nu_{\mu},t)}
{w_{-\mu}(n|n-\nu_{\mu})P(n,t)}
\label{eratetot}
\end{equation}
and
$$\rm \dot{S}_{m}(t)= \frac{1}{2}\sum_{n,\mu}[w_{\mu}(n-\nu_{\mu}|n)
P(n-\nu_{\mu},t)-w_{-\mu}(n|n-\nu_{\mu})P(n,t)]$$
\begin{equation} 
\rm\times\hspace{0.1cm}{\rm ln}\frac{w_{\mu}(n-\nu_{\mu}|n)}{w_{-\mu}
(n|n-\nu_{\mu})}.
\label{eratesur}
\end{equation}

At steady state, $\rm \dot{S}_{sys}=0$ (whether equilibrium or 
NESS). An NESS is characterized by a non-zero total epr 
given by 
$$\rm\dot{S}_{\rm tot}^{\rm (NESS)}
= \sum_{n}\left[w_{1}(n-1|n)P(n-1)-w_{-1}(n|n-1)P(n)]\right.$$
\begin{equation}
\rm \times{\rm ln}\left (\frac{w_{1}(n-1|n)P(n-1)\times w_{-2}(n|n-1)P(n)}
{w_{-1}(n|n-1)P(n)\times w_{2}(n-1|n)P(n-1)}\right).
\label{stotness} 
\end{equation} 
This equation is derived using the circular balance 
condition \cite{vellea} 
$$\rm w_{1}(n-1|n)P(n-1)-w_{-1}(n|n-1)P(n)=$$ 
\begin{equation}
\rm w_{-2}(n|n-1)P(n)-w_{2}(n-1|n)P(n-1).
\label{nesseq1}
\end{equation} 

Now here we consider two limiting situations.
 
(i) It is clear that 
if we do not consider the presence of the product 
species then there will be just two sets of rate 
constants, $\rm k_{1}^{(n)}$ and 
$\rm k_{-1}^{(n)}$. 
Then at the steady state, the balance condition that holds 
is obviously the detailed balance which gives 
$$\rm w_{1}(n-1|n)P(n-1)-w_{-1}(n|n-1)P(n)=0.$$ 
Then from Eq.(\ref{stotness}), we have $\rm\dot{S}_{\rm tot}=0$ at 
the steady state which is reduced now to an equilibrium. 
Now from Eq.(\ref{Kmstpwise}), 
the quantity $\rm X^{(j)}$ under this condition becomes 
$\rm X^{(j)}=\frac{k_{1}^{'(j)}[S]}{k_{-1}^{(j+1)}}$ 
where $\rm \frac{k_{1}^{'(j)}}{k_{-1}^{(j+1)}}$ 
are the stepwise equilibrium (binding) constants. 
So in this limit, theoretically there is no  difference between our 
model and a protein-ligand binding model which generally does not consider 
the product formation. 

(ii) Another interesting point is that 
if we consider the case where $\rm k_{1}^{(n)},k_{-1}^{(n)}\quad>>k_{-2}^{(n)}$ 
(with $\rm k_{2}^{(n)}$ being already considered negligible) then 
this corresponds to the pre-equilibrium limit or simply the 
equilibrium limit. The assumption is valid when fast reversible
 reactions precede 
slower reactions in a reaction network. Now in this situation 
also, the quantity $\rm X^{(j)}$ is defined in terms of the 
stepwise equilibrium (binding) constants. In this context, 
we mention that the original derivation of the enzyme catalysis 
reaction by Michaelis and Menten involved the pre-equilibrium 
assumption with the equilibrium dissociation constant parameter. 
The more general derivation by Briggs and Haldane used the steady state 
approximation and their expression contained the actual 
Michaelis-Menten constant. In our case also we see the same features in the 
quantity, $\rm X^{(j)}$ which is the parameter of our 
model study. In the general nonequilibrium case, 
$\rm X^{(j)}$ is related to the stepwise Michaelis-Menten constant, 
$\rm K_M^{(j)}$ (see Eq.(\ref{Kmstpwise}) with $\rm k_{2}^{(j)}$ 
considered negligible) whereas in the absence of product 
species leading to equilibrium or under the pre-equilibrium 
assumption, $\rm X^{(j)}$ is related to the stepwise equilibrium 
(binding) constant, $\rm \frac{k_{1}^{'(j)}}{k_{-1}^{(j+1)}}.$

{\section{ Numerical simulation of entropy production}}

In this section, we have calculated the medium, system and the total 
entropy production for the spatial and temporal cooperative 
systems. For a 
given initial condition, the oligomeric enzyme system reaches NESS at 
a particular time which depends on the chemiostatic condition, {\it i.e}, 
the value of the constant substrate concentration. 
The initial condition is taken as the fully unbound state of the enzyme 
with all the subunits being vacant {\it i.e.,} 
 $\rm P(n,t=0)=\delta_{n,0}$. This condition leads to zero
system entropy at t=0.  
For the time-dependent 
system entropy production calculation in general, one needs the 
time-dependent solution of the master equation, $\rm P(n,t)$. 
But here the final time in the calculation of the entropy production 
over the time interval (starting at $\rm t=0$ with the specified 
initial condition above) is taken such that by then 
the system reaches the NESS and hence steady state solutions 
are all we need to get the system entropy production 
over the length of the trajectory. 
The total entropy production for 
a single trajectory is calculated over the time 
interval where the determination of the medium entropy production 
requires the detailed information of the path and 
not just the initial and final points. 
We run the simulations in all the cases up to a 
fixed point of time taken to be the same for all the binding mechanisms. 
As the steady state is an NESS (and not an equilibrium), 
total and medium entropy production 
%$\rm \Delta S_{tot}$ and  $\rm \Delta S_{m}$ 
increase linearly with time and 
hence if the final point of time is not the same for 
all the cases, one can not compare the various 
entropy production values for the different cooperative systems.

\subsection{Implementation of the scheme of single trajectory stochastic
 simulation}
 
Along a single  stochastic trajectory the system entropy production  
 can be defined as\cite{seifert4} 
$\rm \textbf{S(t)}=-{\rm ln}\hspace{0.1cm} p(n,t)$, 
where $\rm p(n,t)$ is the solution of the stochastic master equation for 
a given initial condition, $\rm p(n_{0},t_{0})$, taken 
along the specific trajectory $\rm n(t)$. 
Note that, the single trajectory entropy is denoted by (bold) $\textbf{S}$ 
whereas the trajectory-average entropy production (equivalent to ensemble
 average) is denoted by  $\rm S$. Now at the microscopic level, the number 
of occupied sites of the 
oligomeric enzyme becomes a fluctuating quantity due to the random 
occurrence of the different reaction events within the random time interval. 
This develops the concept of different trajectories. 
Here the state of the system can change by any one of the four reactions 
(denoted with index $\mu$) as discussed in Sec.IIC. 

A stochastic trajectory,
 $\rm n(t)$ starting at the state $\rm n_{0}$, jumping at times $\rm t_{j}$ 
from the state $\rm n_{j-1}$ to the state $\rm n_{j}$ and finally ending up at 
$\rm n_{l}$ with $\rm t=t_{l}$ is defined as, 
\begin{equation}
\rm n(t)\equiv {({n_{0}},t_{0})\stackrel{{\nu_{\mu}}^{(1)}}{\rightarrow} 
({n_{1}},t_{1})
\stackrel{{\nu_{\mu}^{(2)}}}{\rightarrow}%({n_{2}},t_{2})\rightarrow
{.....}\rightarrow
({n_{j-1}},t_{j-1})\stackrel{{\nu_{\mu}}^{(j)}}{\rightarrow}({n_{j}},t_{j})
\rightarrow{...}\rightarrow({n_{l-1}},t_{l-1})\stackrel{{\nu_{\mu}}^{(l)}}
{\rightarrow}({n_{l}}},t_{l}).
\label{fordtraj}	
\end{equation}
Here $\rm n_{j}=n_{j-1}+\nu_{\mu}^{(j)}$ where $\nu_{\mu}^{(j)}$ is the 
stoichiometric coefficient of the $\mu$-th reaction along a trajectory 
and $\rm t_{j}= t_{j-1}+\tau_{j}$ 
where 
$\rm \tau_{j}$ is the time interval between two successive jumps. 
During the jump from the $\rm (n_j-1)$ state to the $\rm n_j$ state, 
any one of the four 
reactions will occur (see Eq. (\ref{e2esn}) and Eq. (\ref{e2es})). 
The rate constant of the reaction $\mu$ is denoted
 as $\rm k_{\mu}$. The time interval $\rm \tau_{j}$ between the 
two jumps is a random 
variable following the exponential distribution\cite{gillespie1,gillespie2} 
\begin{equation}
\rm p(\tau_j)=a\hspace{0.2cm} exp(-a\tau_j)
\end{equation}
with
$\rm a=\sum_{\mu=\pm{1}}^{\pm{2}} w(n_j-1;\nu_{\mu}^j)$. 
Here $\rm w(n_{j-1};\nu_{\mu}^{(j)})$ denotes the forward 
transition probability from 
the state $\rm (n_j-1)$ to the state $\rm n_j$ through a reaction 
channel $\rm \mu$ 
with the stoichiometric coefficient $\rm \nu_{\mu}^{(j)}$.

Now a time reversed trajectory can be defined as,
\begin{equation}
\rm n^{R}(t) \equiv (n_{l},t_{l})\stackrel{{-\nu_{\mu}}^{(l)}}{\rightarrow} 
(n_{l-1},t_{l-1})\stackrel{{-\nu_{\mu}}^{(l-1)}}{\rightarrow}{...}
\rightarrow(n_{j},t_{j})\stackrel{{-\nu_{\mu}}^{(j)}}{\rightarrow}
(n_{j-1},t_{j-1}){...}
\rightarrow(n_{1},t_{1})\stackrel{{-\nu_{\mu}}^{(1)}}{\rightarrow}(n_{0},t_{0}).
\label{backtraj}
\end{equation}
This time reversed trajectory is generated due to the occurrence of a 
reaction channel whose state changing vector $\rm {-\nu_{\mu}^{(j)}}$ is 
exactly opposite to 
the state changing vector $\rm {\nu_{\mu}^{(j)}}$ of the forward reaction
 channel.

The time-dependent total entropy production, $\rm \Delta \textbf{S}_{tot}$ 
along a trajectory 
can be split into a system part, $\rm \Delta \textbf{S}_{sys}$ and 
a medium contribution, $\rm \Delta \textbf{S}_{m}$. 
Hence the change of total entropy along a trajectory can be written 
as \cite{seifert4}
\begin{equation}
\rm \Delta \textbf{S}_{tot}=\Delta \textbf{S}_{m}+\Delta \textbf{S}_{sys}
\label{deltasi}
\end{equation}
where
\begin{equation}
\rm \Delta \textbf{S}_{sys}={\rm ln} \frac{p(n_{0},t_{0})}{p(n,t)}
\label{deltas}
\end{equation}
and
\begin{equation}
\rm \Delta \textbf{S}_{m}=\sum_{j}{\rm ln}\frac{w(n_{j-1};\nu_{\mu}^{(j)})}
{w(n_{j};-\nu_{\mu}^{(j)})}.
\label{deltasm}
\end{equation}
Here $\rm w(n_{j-1};\nu_{\mu}^{(j)})$ denotes the forward 
transition probability as already defined.  
Similarly, $\rm w(n_{j};-\nu_{\mu}^{(j)})$ denotes the backward 
transition probability from 
the state $\rm n_j$ to the $\rm (n_j-1)$ state through a 
reaction channel $\rm \mu$ with the exactly 
opposite stoichiometric coefficient $\rm -\nu_{\mu}^{(j)}$. 

\subsection{Cooperative kinetics}

\begin{figure}
\centering
\rotatebox{270}{
\includegraphics[width=8cm,keepaspectratio]{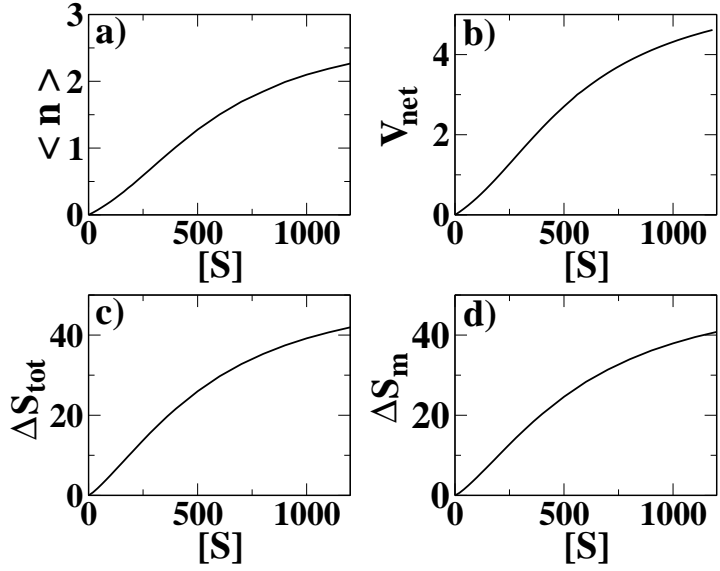}}
\caption{(a) $\rm \langle n \rangle$ and (b) $\rm v_{\rm net}$ for the spatial 
cooperative (sequential) binding as a function of 
substrate concentration, [S] 
(in $\rm \mu M$ unit) at the NESS. 
(c) and (d) exhibit the 
corresponding $\rm \Delta S_{tot}$ and $\rm \Delta S_{m}$ variations with [S]. 
The entropy productions are calculated over a time interval 
that starts with the given initial condition (see text) 
and ends with the system at the NESS.}
\label{fig_2}
\end{figure}

To simulate the spatial cooperativity associated with the 
sequential binding, we have taken the site-independent 
reaction rate constants as $\rm k_{1}^{'}=0.015$ $\rm \mu M^{-1}s^{-1}$ and  
$\rm k_{-1}=7.0 $, 
$\rm k_{-2}=2.0 $, $\rm k_{2}=0.001 $, all in $\rm s^{-1}$. The 
substrate concentration is taken in $\rm \mu M$ unit.
The total number of subunits present in the oligomeric enzyme is taken as 
$n_{\rm T}=3.$ 
We have calculated 
the various entropy productions using the stochastic simulation 
for single trajectories over a time interval starting from the 
initial condition to a final time as mentioned above. 
We have taken $\rm 2 \times 10^{5}$ 
trajectories to get the ensemble average of the entropy production values. 
We have calculated the average binding number, $\langle n \rangle$
 for this case from Eq.($\ref{avgpopspace}$) and the net product 
formation rate by using the formula, 
$\rm v_{\rm net}= \rm k_{-2}\langle n \rangle - k_{2} \langle n_{\rm T}- n 
\rangle$, at the final time where the system 
resides at the NESS. 
We have plotted these quantities as a function of the substrate concentration 
in figure \ref{fig_2}(a) and (b). It is clear from the plots that both the 
quantities grow with a sigmoidal shape as a function of substrate 
concentration indicating positive cooperativity in substrate binding. 
According to Eq.($\ref{avgpopspace}$), this is due to the higher power 
(\rm $>1$) dependence of $\rm \langle n \rangle$ on the factor 
$\rm X$ which is proportional to 
the substrate concentration. As the rate constants are taken as 
site-independent, 
the positive cooperativity generated in the system is inherent in the 
binding mechanism. 
Now we have plotted $\rm \Delta S_{tot}$ and $\rm \Delta S_{m}$, both being 
ensemble averages taken over the $\rm 2 \times 10^{5}$ realizations of the 
trajectories, in 
figure \ref{fig_2}(c) and (d), respectively, against the substrate 
concentration. Interestingly, we find the nature of both the curves 
to be sigmoidal. 

\begin{figure}
\centering
\rotatebox{270}{
\includegraphics[width=8cm,keepaspectratio]{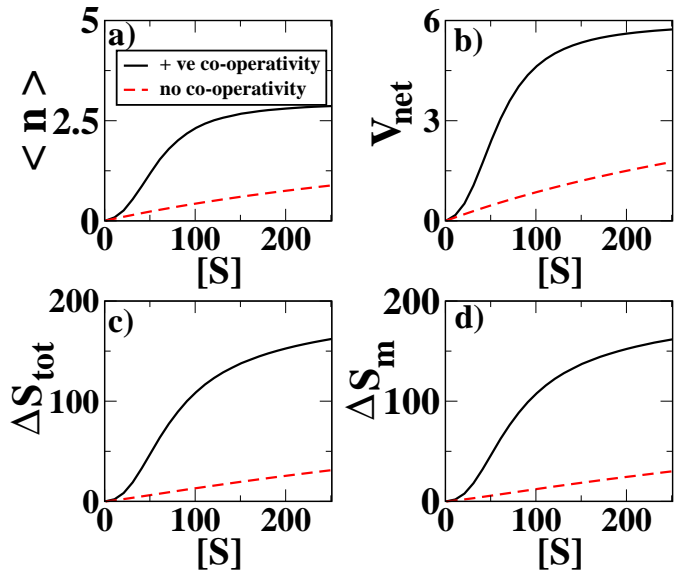}}
\caption{(a) $\rm \langle n \rangle$ and 
(b) $\rm v_{\rm net}$ for the temporally
 cooperative (independent) binding with positive cooperativity against 
substrate concentration, [S] (in $\rm \mu M$ unit) at the NESS. 
(c) and (d) give the corresponding 
$\rm \Delta S_{tot}$ and $\rm \Delta S_{m}$ variations with [S]. 
The entropy productions are calculated over a time interval 
as described in the caption of figure \ref{fig_2}. 
It is evident from the figure that all the curves show a significant 
sigmoidal behavior indicating the positive cooperativity.}
\label{fig_3}
\end{figure}

Next we come to the case of independent substrate binding that can give 
rise to the case of temporal cooperativity with 
site-dependent reaction rate constants. 
To simulate the entropy production for the positive 
cooperative system, we take the rate constants of successive 
substrate binding steps as (see Eq.(\ref{poscop})): 
$\rm k_{1}^{(1)}=f^{(1)}k_{1}^{(0)} $ and 
$\rm k_{1}^{(2)}=f^{(2)}k_{1}^{(0)}$,  
where $\rm k_{1}^{(0)}=\rm k_{1}^{'(0)}[S]$ with 
$\rm k_{1}^{'(0)}=0.015$ $\rm \mu M^{-1}s^{-1}$. 
The set $\rm \{ k_{1}^{'(0)},k_{-1},k_{-2},k_{2}\}$ is called the starting or 
initial rate constants of the cooperative system. 
For the simulation, here we take $\rm f^{(1)}=10$ and $\rm f^{(2)}=100$,
 {\it i.e.}, 
a 10-fold increase in substrate binding rate constants in each step. 
The other rate constants are site-independent and taken to be 
the same as in the case of the 
spatial cooperativity. 
We also calculate the average binding number using Eq.(\ref{navrgind}) and 
the net product formation rate, at the NESS. They are shown in figure 
\ref{fig_3} (a) and (b) along 
with the total and the medium entropy production in  
figure \ref{fig_3}(c) and (d), respectively, 
all as a function of the substrate concentration. 
It is evident from the 
figure that all the curves show a significant sigmoidal behavior 
indicating the positive cooperativity. We have also given the 
corresponding quantities in the case of non-cooperativity in the 
same plot for comparison. The non-cooperative case is simulated with 
site-independent rate constants same as in the case of the 
spatial cooperativity. In this case $\rm \langle n \rangle$ is determined 
using Eq.(\ref{navrgno}). We see that in this case also, the nature of 
variation of $\rm \langle n \rangle$, $\rm v_{net}$, $\rm \Delta S_{m}$ 
and $\rm \Delta S_{tot}$ with 
the substrate concentration is the same, hyperbolic to be specific. 

\begin{figure}
\centering
\rotatebox{270}{
\includegraphics[width=8cm,keepaspectratio]{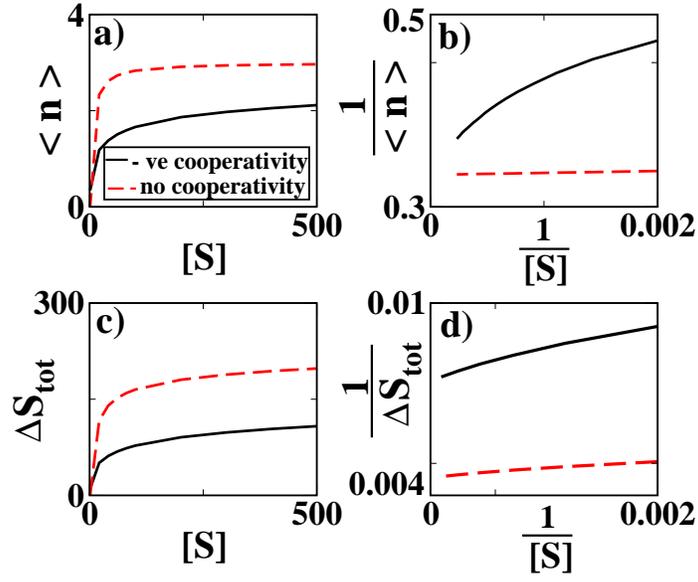}}
\caption{Plots of (a) $ {\rm {\langle n\rangle}}$ vs. [S] 
(in $\rm \mu M$ unit) 
and (b) $ \rm \frac{1}{\rm {\langle n\rangle}}$ 
vs. $\rm \frac{1}{\rm [S]} $ at the NESS. 
(c) $ \rm \Delta S_{tot} $ vs. [S]
 and (d) $ \frac{1}{\rm \Delta S_{tot}} $ versus $\rm \frac{1}{\rm [S]}$
 for negative cooperative (temporal) as well as 
non-cooperative binding. 
The entropy productions are calculated over a time interval 
as described in the caption of figure \ref{fig_2}.}
\label{fig_4}
\end{figure}

Now we come to the last case in this category, {\it i.e.}, the 
negative cooperativity. In this case, the rate constants of the substrate 
binding reaction are taken as (see Eq.(\ref{neco})): 
$\rm k_{1}^{'(0)}=1.5$ $\rm \mu M^{-1}s^{-1}$, 
$\rm k_{1}^{(1)}=f^{(1)}k_{1}^{(0)} $ and 
$\rm k_{1}^{(2)}=f^{(2)}k_{1}^{(0)}$ 
with the values of the factors being $\rm f^{(1)}=0.1$ and $\rm f^{(2)}=0.01$,
 {\it i.e.}, a 10-fold decrease in substrate binding rate constant 
in each step. 
The other rate constants are taken as in the previous cases. 
The value of $\rm k_{1}^{'(0)}$ is taken 
to be $\rm 100$ times greater compared to the cases of spatial and positive 
cooperativity. This is only for the demonstration of the negative 
cooperativity effect on the binding curves of the reaction. 
We have plotted 
$\rm \langle n\rangle$ against substrate concentration 
in figure \ref{fig_4}(a)
 for the negative as well as the non-cooperative case. 
Here for the non-cooperative case also we have taken 
$\rm k_{1}^{'(0)}=1.5$ $\rm \mu M^{-1}s^{-1}$. 
Both the curves show the hyperbolic nature. The two cases are distinguished 
by plotting 
$ \rm \frac{1}{\rm {\langle n\rangle}} $ versus $\rm \frac{1}{\rm [S]} $ 
which is the Lineweaver-Burk plot. 
For non-cooperative enzyme, this plot gives a 
straight line whereas the curve for the negative cooperative binding 
starts at a higher value on the y-axis and becomes nonlinear when it 
comes close to the curve
 of the non-cooperative system at high substrate concentration. 
This feature is shown in figure \ref{fig_4}(b). 
Now we have plotted similar curves for $ \rm \Delta S_{tot} $ in 
figure \ref{fig_4}(c) and (d). One can see the same hyperbolic nature 
in the plot of $ \rm \Delta S_{tot} $ versus substrate concentration
 (figure \ref{fig_4}(c)) 
for both the cases and the nonlinearity in the plot of 
$ \frac{1}{\rm \Delta S_{tot}} $ versus 
$\rm \frac{1}{\rm [S]} $ at high substrate concentration for the negative 
cooperativity (figure\ref{fig_4}(d)). 
So from the above discussion and the plots, we conclude that 
the familiar indications of the cooperative behavior in substrate binding, 
given in terms of the nature of variation of the average binding number and 
the net velocity of the reaction 
as a function of the substrate concentration, are all reflected in 
the same manner in the corresponding variation of the total as well as 
the medium entropy production.

\begin{figure}
\centering
\rotatebox{270}{
\includegraphics[width=8cm,keepaspectratio]{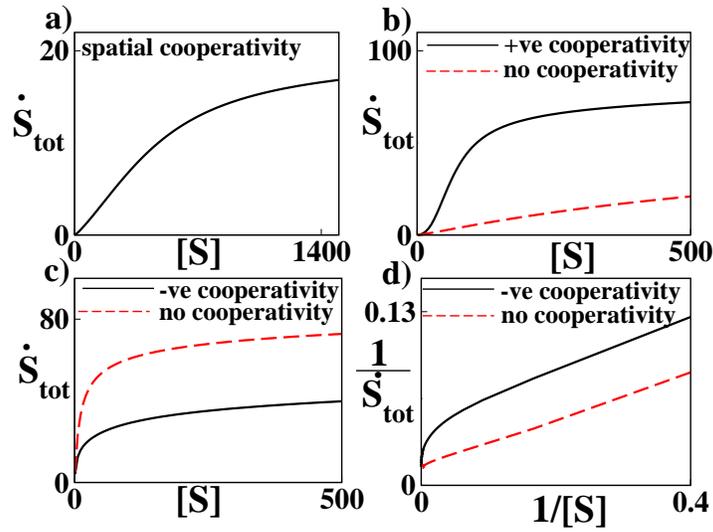}}
\caption{(a) Plot of $ \rm \dot S_{tot} $ against substrate concentration, 
[S] (in $\rm \mu M$ unit) for spatial cooperativity. 
In (b) and (c), the same quantity is plotted for positive and negative 
cooperative cases, respectively. The non-cooperative case is also 
shown for comparison. 
(d) Plot of $\rm \frac{1}{ \rm \dot S_{tot}}$ 
vs. $\rm \frac{1}{[S]}$, which is a Lineweaver-Burk type plot, 
for negative and non-cooperative cases.}
\label{fig_5}
\end{figure}

We have also calculated the total entropy production rate, 
$\rm \dot S_{tot}$ at the NESS 
using Eq.(\ref{stotness}) 
for all the cases of cooperativity . Here we have taken 
the same set of rate constants as we have already considered 
to calculate the various entropy productions. 
The variations of $\rm \dot S_{tot}$ with substrate 
concentration, [S] for different binding schemes are 
shown in figure \ref{fig_5}. It is evident from the 
figure that the features of cooperative binding are 
also reflected in a similar fashion on the variation 
of $\rm \dot S_{tot}$ with substrate concentration. 

\subsection{System entropy production and binding characteristics }

\begin{figure}
\centering
\rotatebox{270}{
\includegraphics[width=8cm,keepaspectratio]{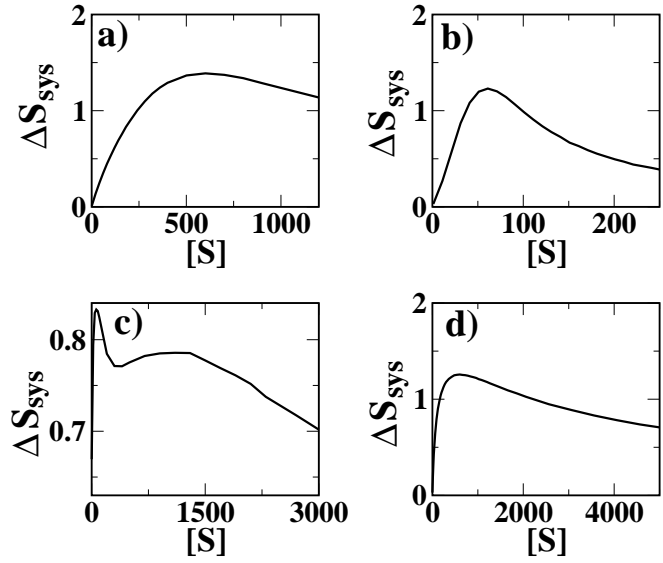}}
\caption{Plot of $\rm \Delta S_{sys}$ against substrate concentration, 
[S] (in $\rm \mu M$ unit) for (a) spatial cooperative binding, 
(b) positive (temporal) 
cooperative binding, (c) negative (temporal) cooperative binding  
and (d) non-cooperative binding. In all the cases, the final time of 
observation is the same, set as such that the system reaches the NESS.}
\label{fig_6}
\end{figure}

The ensemble or trajectory average system entropy production 
over the time interval can be written as 
\begin{equation}
\rm \Delta S_{sys}= S_{sys}^{final}-S_{sys}^{initial}=
 -\sum_{n=0}^{n_T} P^{ss}(n)lnP^{ss}(n),
\label{sysent}
\end{equation}
where the initial condition (time $\rm t=0$) of 
the fully unbound enzyme gives $\rm S_{sys}^{initial}=0$ and
 the final state of the system is an NESS characterized by 
the distribution  $\rm P^{ss}(n)$. 
We have plotted the ensemble average system entropy production, 
$\rm \Delta S_{sys}$ as a function 
of the substrate concentration in figure \ref{fig_6} for all the cases.
In figure \ref{fig_6}(a), $\rm \Delta S_{sys}$ is plotted for spatial 
cooperativity and 
in figure \ref{fig_6}(b-d) it is shown for the positive, 
negative and non-cooperative cases, respectively 
which belong to the class of temporal cooperativity. 
The first thing evident from the plots is that $\rm \Delta S_{sys}$ 
passes through a global maximum for all the cases and in the 
case of negative cooperativity, 
there is also a local maximum with the parameters of our system. 

\begin{figure}
\centering
\rotatebox{270}{
\includegraphics[width=8cm,keepaspectratio]{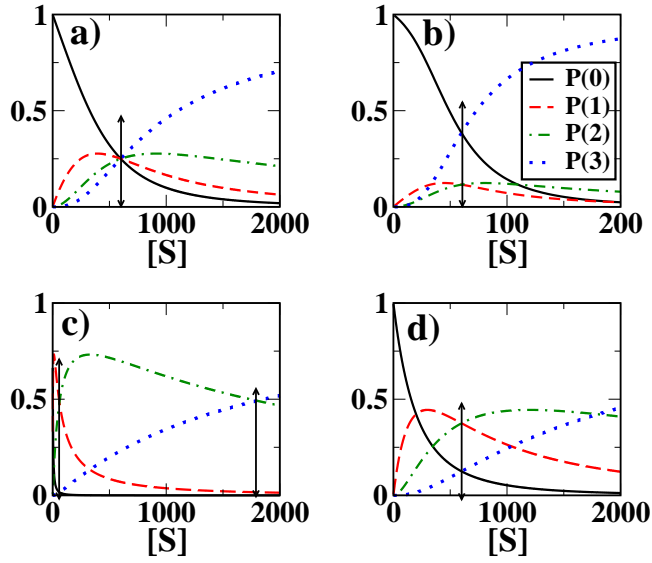}}
\caption{Plot of the steady state distributions, $\rm P^{ss}(n)$ against 
[S] (in $\rm \mu M$ unit)  
for (a) spatial cooperative binding, (b) positive (temporal) 
cooperative binding, (c) negative (temporal) cooperative binding  and 
(d) non-cooperative binding. In the plots, the `ss' superscript
is dropped for simplicity. The arrows indicate the curve crossing points.}
\label{fig_7}
\end{figure}

We have plotted $\rm P^{ss}(n)$ as a function of the substrate concentration 
in figure \ref{fig_7}(a-d) with the steady state (`$\rm ss$') superscript 
being dropped for simplicity. Figure \ref{fig_7}(a) shows the curves for 
spatial cooperativity. 
We can see that they all cross almost exactly at the same point, 
$\rm [S] \sim 600\hspace{0.1cm} \mu M$ giving rise to the maximum 
in $ \rm \Delta S_{sys} $ 
for spatial cooperativity at this point (see figure \ref{fig_6}(a)). 
In figure \ref{fig_7}(b), we have shown the curves 
for the positively cooperative system. At $\rm [S]\sim 60\hspace{0.1cm}\mu M$, 
the curves cross in a pairwise fashion; curves of 
$ \rm P(0) $ and  $ \rm P(3) $ cross each other at this point as well 
as curves for $ \rm P(1) $ and  $ \rm P(2) $. 
This particular substrate concentration corresponds to the maximum of 
$ \rm \Delta S_{sys} $ in this case (see figure \ref{fig_6}(b)).

The case of negative cooperativity requires a bit more 
attention. There is again pairwise curve 
crossing of the two sets of probabilities same as in the case 
of positive cooperativity at the same substrate concentration 
shown in figure \ref{fig_7}(c). This gives 
rise to the global maximum in the curve of $ \rm \Delta S_{sys} $ for 
this type of binding shown in figure \ref{fig_6}(c). 
The local maximum can be explained as 
follows. Unlike the plots in figure \ref{fig_7}(a) and (b), the probability 
curves $ \rm P(2) $ and  $ \rm P(3) $ remain at significant values 
over the substrate range studied and the dominance of these two 
probability curves in figure \ref{fig_7}(c)
 (actually when $ \rm P(2) $ and 
$ \rm P(3) $ cross, they are close to $\rm 0.5$ at 
$\rm [S] \sim 1800 \hspace{0.1cm}\mu M$) over a 
large substrate range 
gives rise to an increase  of $ \rm \Delta S_{sys} $, albeit slow.  
Finally we come to the 
case of non-cooperativity in figure \ref{fig_7}(d) where again there is the 
pairwise crossing of the same set of probabilities as in 
figure \ref{fig_7}(b) but 
at $\rm [S] \sim 600 \hspace{0.1cm}\mu M$ that again 
gives rise to the maximum of 
$ \rm \Delta S_{sys} $ shown in figure \ref{fig_6}(d). 
In this case too, there are more than one 
dominating probability curves before and after the 
pairwise crossing over similar substrate range as 
in figure \ref{fig_7}(c). But the $ \rm \Delta S_{sys} $ in this case 
shows a slow but steady decrease with substrate 
concentration after passing through the maximum 
without any unusual behavior. This may be due to the 
fact that here at least three of the four probabilities 
are significant (with comparable values) over a large substrate range 
and so they do not cross the value of $\rm 0.5$ in this range unlike the case 
in figure \ref{fig_7}(c). It is clear that arbitrary variation of the rate 
constants of the system in each binding step can make life more complicated
 and then the maxima in the $ \rm \Delta S_{sys} $ curve may or 
may not be associated with the binding probability curve crossings. 

We can justify the curve crossings, whether they all cross or 
cross pairwise at a particular substrate concentration, by inspecting 
the expressions of the steady state 
probability distributions. 
We see from the steady state distribution for the spatial 
cooperativity, Eq.(\ref{geometricdistribution}), that if 
one of the probabilities, say $\rm P(0)$ is approximately equal 
to any other probability, 
say $\rm P(3)$, then $\rm X\sim 1$ (but obviously not 
exactly equal to 1) and this automatically leads to the 
near equality of all the probabilities at this value of $\rm X$. 
This is true for all the probabilities and hence in this 
case the probabilities can only cross simultaneously 
at $\rm X\sim 1$. As the probabilities are equal at this point 
which corresponds to $\rm [S]\sim 600 \hspace{0.1cm}\mu M$, this obviously 
gives the maximum system entropy production in this case. 
Now we take the steady state distribution of the 
non-cooperative case, Eq.(\ref{binom}). It can be 
easily seen that here only $\rm P(0)=P(3)$ leads
to the equality $\rm P(1)=P(2)$ at $X=1$  giving the 
maximum of $ \rm \Delta S_{sys} $ again at
 $\rm [S]\sim 600 \hspace{0.1cm}\mu M$. 
So in the context of the system entropy production 
 the spatial cooperative system shows 
some similarity with the non-cooperative system. 
With the same set of site-independent rate constants, the 
spatial cooperative system is also associated with larger system 
entropy production  compared to that of 
the non-cooperative case. This is because all the binding probabilities 
become equal for the spatial cooperativity whereas they 
become equal pairwise for non-cooperative binding. 

The cases of positive and negative cooperativity belonging 
to the class of temporal cooperativity 
are a bit complicated. We have considered a 10-fold 
increase of the substrate binding rate constant for each 
successive binding in the case of positive cooperativity 
whereas a 10-fold decrease in the corresponding rate constant 
for each successive binding for negative cooperativity. 
This symmetry ensures that in both the cases only $\rm P(0)=P(3)$ leads
to the equality $\rm P(1)=P(2)$ at $\rm X^{(1)}=1$. 
This can be easily proved from Eq.(\ref{stdisttemp}). 
But if the rise or fall of the value of the substrate binding rate 
 constant in each successive step of binding is not by the same factor, 
then the pairwise equality of the binding probabilities is 
not possible at a given substrate concentration.

\section{ Measure of cooperativity}

Here we have discussed on the determination of the Hill coefficient 
from the master equation corresponding to the different binding schemes. 
We have also introduced an index of cooperativity in terms of 
the stochastic system entropy 
associated with 
the fully bound state of the cooperative and non-cooperative cases. 
 We have analyzed its connection to the Hill coefficient 
using some relevant experimental data which gives a realistic 
application of the proposed scheme of measurement of cooperativity.

\subsection{Hill coefficient}

In the traditional enzymology, the characterization of cooperativity
is carried out by measuring the Hill coefficient \cite{palmer}. For positive
and negative cooperative cases, the Hill coefficient becomes
greater than or less than one, respectively, whereas 
the non-cooperative case is characterized with Hill coefficient
equal to one.
Experimentally it is obtained by determining the fractional 
saturation, $\theta$($=\rm \langle n\rangle/n_T$) at various substrate 
concentrations $\rm [S]$, constructing the 
Hill plot ($\rm ln(\frac{\theta}{1-\theta})$ vs. $\rm ln[S]$) 
and then finding the slope at 
the half-saturation point, $\rm \theta=0.5$ or at a point 
where the slope deviates maximum from unity. 
On the other hand, Hill coefficient is theoretically defined as
 the ratio of the 
variances of the binding number of the cooperative and non-cooperative 
cases at the half-saturation point with the non-cooperative binding case
 following the binomial distribution\cite{abeliovich,wyman}. 

We briefly mention the features of the 
Hill plot for the model binding schemes studied here. 
The slope of the Hill plot is generally given by \cite{abeliovich}
\begin{equation}
\rm H=\frac{[S](d\theta/d[S])}{\theta(1-\theta)}.
\label{slope}
\end{equation}
For temporal cooperativity, the fractional saturation 
can be written as (see Eq.(\ref{navrgind}))
\begin{equation}
\rm \theta_{temp}=\frac{\sum_n nB_n[S]^n}{n_T\sum_n B_n[S]^n},
\end{equation}
where $\rm B_n=\binom{n_T}{n}\prod_{j=0}^{n-1}(K_M^{(j)})^{-1}$ 
with $\rm B_0=1.$ 
Then one gets 
\begin{equation}
\rm H_{temp}=\frac{\langle n^2\rangle - \langle n\rangle^2}{n_T\theta(1-\theta)}=\frac{\sigma^2_{temp}}{\sigma^2_{bino}},
\label{Htemp}
\end{equation}
where $\rm \sigma^2_{temp}$ and $\rm \sigma^2_{bino}$ are the 
variances of the binding numbers of the temporal 
and non-cooperative cases, respectively. The Hill coefficient, 
$\rm n_H$ is given at the half-saturation point as\cite{abeliovich} 
$\rm n_H=\frac{4\sigma^2_{temp}}{n_T}$. 
Similar expressions hold for the spatial cooperative 
binding. 
We have already mentioned in Sec.IIB that if all the 
rate constants of the independent binding scheme are site-independent, 
then the binding is non-cooperative with binomial 
distribution of the binding probability. 
Here we discuss the corresponding scenario for 
the sequential binding (leading to spatial cooperativity) 
to be non-cooperative 
in terms of the variance of the binding number. 
In the case of spatial cooperativity, the variance of the 
binding number, $\rm \sigma^2_{sp}$ is given by 
\begin{equation}
\rm \sigma^2_{sp}=\Big(\frac{1+X}{1-X}\Big)\langle n\rangle -
 \langle n\rangle^2 -
\frac{n_T(n_T+1)X^{n_T+1}}{1-X^{n_T+1}},
\label{varsp}
\end{equation}
where $\rm X$ and $\langle n\rangle$ are as given in
 Eq.(\ref{geometricdistribution}) and 
Eq.(\ref{avgpopspace}). 
Now for $\rm n_T=1$, this reduces to 
\begin{equation}
\rm \sigma^2_{sp}=\frac{X}{(1+X)^2}=\rm \sigma^2_{bino}.
\end{equation}
and then the slope of the Hill plot becomes 
$ \rm H_{sp}=\frac{\sigma^2_{sp}}{\sigma^2_{bino}}=1$ for 
any substrate concentration. 
So for sequential binding, 
the cooperative behavior is absent only if the 
enzyme is monomeric.

\subsection{Cooperativity index}

Here we introduce an index of cooperativity. 
First we build up the concept from binding probabilities 
and then 
demonstrate how this index can 
indicate the nature of the cooperativity. 
For positive cooperative 
binding, one expects that full occupancy of the enzyme 
is more probable compared to the case of non-cooperativity. 
Similarly, for negative 
cooperativity, the full occupancy of the enzyme is less probable. 
Now, if the probability of an event-n is $\rm p_n$, 
then the associated surprisal is given by $\rm -ln(p_n)$ and 
more probable the event, the less is its surprisal. 
So the ratio of the surprisals, associated with the probability 
of the system to remain in a fully occupied state without 
and with cooperativity at NESS, 
should be greater than $\rm 1$ 
for positive cooperativity and less than $\rm 1$ for negative cooperativity. 
Hence we define the index of cooperativity, denoted by $\rm C$ 
in terms 
of the ratio of the surprisals, associated with the probability 
of the system to remain in a fully occupied state without 
and with cooperativity at NESS as 
\begin{equation}
\rm C=\frac{-ln(P^{(bino)}(n_T))}{-ln(Q(n_T))}
\label{surp}
\end{equation}
where the binomial distribution, $\rm P^{(bino)}$ is the reference 
corresponding to the non-cooperative case and 
the distribution $\rm Q$ corresponds to the cooperative binding case. 
The rate constants of the reference non-cooperative system 
(binomial) must be the same as those of the starting or initial rate 
constants of the cooperative system for the 
comparison to be valid. The relation is then independent 
of the actual value of 
the (constant) substrate concentration. 
We point out that the surprisal is equivalent to the single trajectory 
stochastic system entropy\cite{seifert4,esposito} 
(associated with the fully occupied state). So the index, C is
truly an entropic estimate of cooperativity at the microscopic
level.

Based on the above argument, next we theoretically analyze the 
probability to remain in a fully 
occupied state for different cooperative systems and 
compare that with the non-cooperative case to formulate 
the criteria of cooperativity in terms of C. 
For spatial cooperativity, 
the ratio of its steady state distribution 
(Eq.(\ref{geometricdistribution})) 
and the reference binomial distribution (Eq.(\ref{binom})) 
for $\rm n=n_T$ is given by 
\[\rm R_{sp}=\frac{P_{sp}^{ss}(n_T)}{P^{(bino)}(n_T)}\]
\begin{equation}
\rm =1+\frac{\Big[\Big(\binom{n_T}{1}-\binom{n_T}{0}\Big)X+
\Big(\binom{n_T}{2}-\binom{n_T}{1}\Big)X^2+\ldots+
\Big(\binom{n_T}{n_T}-\binom{n_T}{n_T-1}\Big)X^{n_T}\Big]}
{(1-X^{n_T +1})}.
\label{Rsp}
\end{equation}
>From the above expression it is clear that for all values of $\rm X$, 
either greater than or less than 1, the quantity $\rm R_{sp}$ is greater
 than 1 indicating positive cooperativity. This will lead to the 
condition $\rm C > 1$ for the case of spatial cooperativity for 
any substrate concentration. Only in the case of monomeric enzyme,
($\rm n_{T}=1$),  the system will be non-cooperative with  
$\rm R_{sp}=C=1$ as already discussed in terms of variances at 
the end of Sec.IVA.

In the case of temporal cooperativity, the corresponding 
ratio, $\rm R_{temp}$ is given using Eq.(\ref{stdisttemp}) and 
Eq.(\ref{binom}) at $\rm n=n_T$ as  
$$\rm R_{temp}=\frac{P_{temp}^{ss}(n_T)}{P^{(bino)}(n_T)}$$
$$\rm =\Bigg[\frac{\frac{X^{(0)}X^{(1)}{...}X^{(n_T-1)}}
{[1+n_T X^{(0)}+\ldots+(X^{(0)}X^{(1)}\ldots X^{(n_T-1)})]}}
{\frac{X^{n_T}}{(1+X)^{n_T}}}\Bigg]$$
\begin{equation}
\rm =\Bigg[\frac{\frac{(X^{(0)})^{n_T}f^{(n_T-1)}}
{[1+n_T X^{(0)}+\ldots+(X^{(0)})^{n_T}f^{(n_T-1)}]}}
{\frac{(X^{(0)})^{n_T}}{[1+n_T X^{(0)}+\ldots+(X^{(0)})^{n_T}]}}\Bigg],
\end{equation}
with $\rm X^{(0)}=X$. Now both $\rm P_{temp}^{ss}(n_T)$ and 
$\rm P^{(bino)}(n_T)$ tend to 1 at large $\rm X^{(0)}$ {\it i.e.} 
large substrate concentration. But it is clear that for 
positive cooperative binding with $\rm f>1$, 
the last term in the denominator of $\rm P_{temp}^{ss}(n_T)$ 
dominates the previous terms more readily compared to the 
case of $\rm P^{(bino)}(n_T)$. Hence at a particular substrate 
concentration, $\rm P_{temp}^{ss}(n_T)$ is closer to 1 
compared to $\rm P^{(bino)}(n_T)$ and so $\rm R_{temp}$ is 
greater than 1. For negative cooperativity with $\rm f<1$, 
the situation is obviously reverse and $\rm R_{temp}$ is less than 1. 
Therefore, in the light of the above discussions and Eq.(\ref{surp}), 
we  write down the condition of cooperativity in terms of $\rm C$ as 
\begin{equation}
\rm C  \left\{ \begin{array}{ll}
>1, & \textrm{positive cooperativity}\\
=1, & \textrm{no cooperativity}\\
<1, & \textrm{negative cooperativity}.
\end{array} \right.
\label{Mratio}
\end{equation}
This is the same criteria of cooperativity as given in terms of 
the Hill coefficient. 
To find out the Hill coefficient,
{\it i.e.,} the variances theoretically, it is necessary to know
 the probability distribution of the corresponding positive 
and negative cooperativity cases, respectively.
Now our measure of cooperativity, the index C, is also related to the 
probability distributions; but it is defined in terms of the 
ratio of 
a specific term of the distributions, namely the probability 
of the fully occupied state. So apparently there is no straightforward 
connection between the Hill coefficient and C. 
The Hill coefficient is the slope of the binding curve 
at a particular substrate concentration corresponding to 
the half-saturation point whereas the index C is defined 
independent of the substrate concentration and the characterization  
of cooperativity in terms of C is valid 
at any substrate concentration.

\begin{figure}
\centering
\rotatebox{270}{
\includegraphics[width=8cm,keepaspectratio]{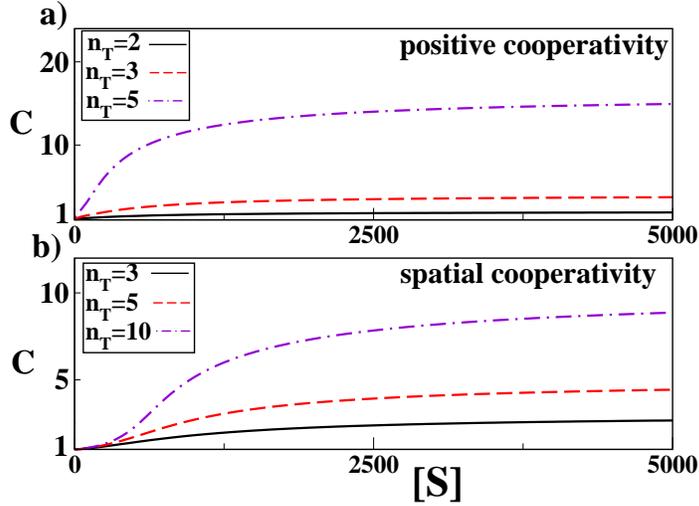}}
\caption{Plot of the cooperativity index, C against substrate 
concentration, [S] (in $\rm \mu M $ unit) for 
different values of the number of subunits of the oligomeric enzyme,
 $\rm n_T$ in the case of (a) positive (temporal) cooperativity and 
(b) spatial cooperativity.}
\label{fig_9}
\end{figure}

We have plotted the quantity, C in figure (\ref{fig_9}) for 
positive cooperative system (independent binding) 
and also for the spatial cooperative binding for different values 
of $\rm n_T$ as a function of substrate concentration. 
For the positive cooperativity case, the substrate binding rate
constants, $\rm k_{1}^{(n)}$ increase by a factor of $2$ in
each step.
The value of C grows with substrate concentration, starting just above 
$1.0$ and finally saturates. One can see from Eq.(\ref{surp}),  
that the limiting value of C (obtained at high substrate concentration)  
in case of spatial cooperativity is 
$\rm n_T$ whereas for temporal cooperativity it is given by 
$\rm f^{(n_{T}-1)}$ where $\rm f^{(n_{T}-1)}=
\frac{k_{1}^{(n_{T}-1)}}{k_{1}^{(0)}}$. These results are discussed in
detail in the appendix.  
Here we specifically mention the case of $\rm n_{T}=5$ for the 
positive cooperativity where the limiting value of $\rm C$ 
is  $\rm f^{(4)}=
\frac{k_{1}^{(4)}}{k_{1}^{(0)}}= 2^{4}=16$. It is evident from figure
(\ref{fig_9})(a) that this is indeed the case.

%%%%%%%%%%%%%%%%%%%%%%%%%%%%%%%%%%%%%%%%%%%%%%%%%%%%%%%%%%%%%%%%

\subsection{Characterization of cooperativity: a case study with 
stepwise Aspartate receptor binding}

Although apparently there is no straightforward 
connection between the Hill coefficient, 
$\rm n_H$ and C, first of all 
it is clear that the well-known criteria of cooperativity 
in terms of $\rm n_H$ is exactly the criteria we have 
given in terms of the cooperativity index, C in Eq.(\ref{Mratio}). 
Both the measures are equal to $1$ in the absence of cooperativity whereas 
for cooperative binding, the criteria are the same 
although the actual values of $\rm n_H$ and C will be 
generally different. 
Here we will try to illustrate this point 
using some experimental data from the work of Kolodziej {\it et al.} 
\cite{tan} regarding 
the production of positive, negative as well as non-cooperativity by 
mutations at a serine 68 
residue located at the subunit interface in the dimeric aspartate receptor of 
{\it Salmonella typhimurium}. Due to unavailability of experimental 
data of the stepwise Michaelis-Menten constants, $\rm K_M^{(j)}$, 
we use the 
stepwise binding constants reported in their study in the place of 
$(\rm K_M^{(j)})^{-1}$ of the independent binding model with 
$\rm n_T=2$ and $\rm j=0,1$. 
Now the parameter $\rm X^{(j)}$ in our study is related to
 $\rm K_M^{(j)}$ in the general non-equilibrium condition and 
reduces to stepwise equilibrium (binding) constants under the 
conditions already discussed at the end of Sec.IIC. 
For experimental testability of C at NESS, one needs  
the stepwise Michaelis-Menten constants, $\rm K_M^{(j)}.$ 
We choose the 
independent binding model as the experimental result 
reports both positive and negative cooperativity. 
We calculate the fractional saturation $\theta$ as a 
function of substrate concentration, $\rm [S]$ using Eq.(\ref{navrgind}) 
and find out the Hill coefficient, $\rm n_H$ at the 
half-saturation point ($\theta=0.5$). 
Then we determine the cooperativity index, C 
at the substrate concentration where $\theta=0.5$. The results 
are given in Table. 1. The Hill coefficients derived by us for 
different cases tally very well with the experimental data\cite{tan}.  
The cooperativity index, C detects the presence and 
absence of cooperativity  
successfully. Also the extent or degree of positive or 
negative cooperative behavior is equally well characterized by 
the index, C. This can be seen by comparing the values of 
$\rm n_H$ and C for the cases of serine and cysteine 
showing negative cooperativity as well as for 
threonine and isoleucine showing positive cooperativity. 

%*************************************************************************
\begin{table}[h!]
\caption{The stepwise Aspartate binding constants, $\rm K_1^{'}$ and 
$\rm K_2^{'}$ (in $\rm \mu M^{-1}$) for different 
amino acid residues at position 68 of Aspartate receptor taken from the 
experimental study of Kolodziej {\it et al}\cite{tan}.
 Here we have taken the values of the inverse of the stepwise 
Michaelis-Menten constants, $\rm K_M^{(j)}$ in our model to be equal to the 
binding constants. The values of the Hill coefficient, $\rm n_H$  in the 
parentheses are from the experimental work, given for comparison 
with the values determined here. 
The cooperativity index, C characterizes the cooperative behavior 
successfully as can be seen by comparing it with $\rm n_H$. }
\begin{tabular}{lccccc}
\hline
\hline
Amino acid & \quad $\frac{1}{\rm K_M^{(0)}}(=K_1^{'})$ & \quad $\frac{1}{\rm K_M^{(1)}}(=K_2^{'})$ & \quad$\rm n_H$ & \qquad C\\
\hline
serine & \qquad 0.7 & \quad 0.2 & \quad 0.7(0.7) & \quad 0.491\\
cysteine & \qquad 0.5 & \quad 0.2 & \quad 0.776(0.8) & \quad 0.598\\
threonine & \qquad 0.4 & \quad 0.9 & \quad 1.197(1.2) & \quad 1.519\\
isoleucine & \qquad 0.4 & \quad 2.8 & \quad 1.446(1.4) & \quad 2.558\\
aspartate & \qquad 0.1 & \quad 0.1 & \quad 1.0(1.0) & \quad 1.0\\
\hline
\hline
\end{tabular}
\label{tab1}
\end{table}
%******************************************************************************

The cooperativity index, C is related to the probability of fully bound 
state of the single enzyme. So another possibility of 
experimentally determining C, apart from the measurement 
of the stepwise Michaelis-Menten constants, 
will be to detect this fully bound state by electrical or optical means in a
 single molecule experiment and then to fit the resulting probability 
with some model distribution.

\section{Conclusion} 

We have classified the cooperative substrate binding phenomena of 
a single oligomeric enzyme on the basis of the binding mechanism 
and the nature of the substrate-bound states of the system 
in a chemiostatic condition. 
Both the binding mechanisms are modelled 
in terms of master equation. 
The sequential binding of the substrate molecules leads to 
spatial cooperativity whereas the independent 
binding scheme leads to temporal cooperativity. 
We have determined the various entropy productions due 
to the enzyme kinetics over a time interval where 
at the final point of time the system is in a 
nonequilibrium steady 
state (NESS) that can be arbitrarily far away from equilibrium. 
We have used kinetic Monte Carlo 
simulation 
algorithm applied on a single trajectory basis to 
calculate the entropy production. 
In this context, the interesting finding is that 
the total as well as the medium entropy 
production show the same diagnostic signatures for 
detecting the cooperativity as is well known 
in terms of the average binding number or the 
net velocity of the reaction. More specifically, 
$\rm \Delta S_{tot}$ as well as $\rm \Delta S_{m}$ 
for positive cooperative kinetics show sigmoidal variation as a function 
of substrate concentration whether the class being 
spatial or temporal. They also show the non-linearity 
in the inverse plot of Lineweaver-Burk type demonstrating 
the case of negative cooperativity. 
The signs of cooperative behavior is also reflected in a 
similar fashion on the 
variation of the total entropy production rate (epr) 
with substrate concentration determined at the NESS for 
different binding schemes. 
That the features of 
cooperativity are reflected similarly on the variations of 
both the total epr 
at the NESS and the total (and medium) entropy production over 
a time interval 
up to the NESS is a highly interesting fact and gives 
deep insight on the role of the binding mechanism in governing 
the total entropy production in a general non-equilibrium setup.

We have thoroughly analyzed the system 
entropy production for all the cases in terms 
of the steady state binding probability distributions. 
For a spatial and a non-cooperative system, 
the maximum value of the system entropy production due to the 
nonequilibrium processes in the reaction 
appears at the same substrate concentration 
with the value of the entropy production 
being greater for the spatial cooperativity. 
We have explained this in terms of the different 
binding probability curve-crossings that 
helps to understand how the binding characteristics 
affect the entropy production of the system, {\it i.e,} 
the single oligomeric enzyme. Similarly, 
the distinct features of the evolution 
of system entropy production for the positive and negative 
cooperative binding give valuable insights on 
its connection to the binding mechanism. 

We have introduced an index of cooperativity, C defined as the 
ratio of the surprisal or equivalently, the stochastic system entropy 
associated with 
the fully bound state of the cooperative and non-cooperative cases. 
The criteria of cooperativity in terms of C is identical to that of the 
Hill coefficient. We have analyzed its connection to the Hill coefficient 
using some relevant experimental data. 
This index is truly an entropic estimate of cooperativity and 
gives a microscopic insight on the cooperative binding of substrate on a 
single oligomeric enzyme instead of realising cooperativity in terms of
macroscopic reaction rate.

\noindent
{\bf Acknowledgement} : K.B. acknowledges the Council of Scientific 
and Industrial Research (C.S.I.R.), India for the partial financial support 
as a Senior Research Fellow.

\appendix*

\section{Estimate of the limiting value of the cooperativity index, C for 
various cooperative binding}

Here the limiting value of the cooperativity index, C for 
the spatial and temporal cooperative binding are determined 
at high substrate concentration. 
The limiting value of the cooperativity index, C for the spatial 
cooperativity is calculated from Eq.(\ref{surp}) 
by using 
the steady state probability distribution function of spatial  
cooperativity, $\rm P_{sp}^{ss}(n)$ (Eq.(\ref{geometricdistribution})) 
and that of no cooperativity, $\rm P_{bino}^{ss}(n)$ (Eq.(\ref{binom}))
at $\rm n= n_{T}$. The expression of C then becomes 

\begin{equation}
\rm C=\frac{-ln \left[(\frac{X}{1+X})^{n_{T}}\right]}
{-ln\left[\frac{X^{n_{T}}(1-X)}{1-X^{(n_{T}+1)}}\right]}.
\label{csequen}
\end{equation}

At high substrate concentration, with $\rm X>>1$, the 
above equation can be written as 
\begin{equation}
\rm C=\frac{n_{T} ln (1+\frac{1}{X})}{-ln (1-\frac{1}{X})}.
\label{shotseq}
\end{equation}

Now expanding the log terms in the Eq. (\ref{shotseq}) and 
neglecting the higher order terms, we finally obtain
\begin{equation}
\rm C=n_{T}.
\end{equation}
Therefore, the limiting value of C in the case of spatial 
cooperativity, obtained at high substrate 
concentration, is equal to the 
total number of sub-units of the oligomeric enzyme.

%\section{Estimate of C for the temporal cooperativity}

In a similar fashion, the limiting value of C can be calculated for the 
temporal cooperativity from Eq.(\ref{surp}) by using 
the steady state probability distribution function of temporal 
cooperativity, $\rm P_{temp}^{ss}(n)$ (Eq.(\ref{stdisttemp})) 
and that of no cooperativity, $\rm P_{bino}^{ss}(n)$ (Eq.(\ref{binom}))
at $\rm n= n_{T}$. 
At this value the distribution, $\rm P_{temp}^{ss}(n)$ 
 can be written as 
\begin{equation}
\rm P_{temp}^{ss}(n_{T})=\frac{X^{(0)}X^{(1)}{...}X^{(n_T-1)}}
{[1+
n_T X^{(0)}+{...}+
(X^{(0)}X^{(1)}{...}X^{(n_T-1)})]}.
\end{equation}
%where $\rm n_{T}=3$.
Here $\rm X^{(j)}\approx f^{(j)}X^{(0)}$ with $\rm j=0,...,(n_T-1)$. 
This follows from  the definition 
of $\rm X^{(j)}$ (see Eq.(\ref{stdisttemp}) and Eq.(\ref{stdistrispace})) 
with the small value of $\rm k_{-2}$ taken in this study. 
Now at high substrate concentration with $\rm X^{(0)}>>1$, 
the above equation can be written as
\begin{equation}
\rm P_{temp}^{ss}(n_{T})=\frac{1}{[1+\frac{n_T}{X^{(0)}f^{(n_T-1)}}]}.
\end{equation}

Now, by using the value of $ \rm P_{temp}^{ss}(n_{T})$ and 
$\rm P_{bino}^{ss}(n_{T})$ into the Eq.(\ref{surp}) at high 
substrate concentration, we obtain
\begin{equation}
\rm C=\frac{n_{T}ln (1+\frac{1}{X})}{ln[1+\frac{n_T}{X^{(0)}f^{(n_T-1)}}]}.
\end{equation}
For the comparative study of the temporal and non-cooperative cases, 
the starting value, $\rm X^{(0)}$ is taken equal to  $\rm X$. 
Then expanding the log terms in the above equation and 
neglecting the higher order terms, we finally obtain the limiting 
value of C for temporal cooperativity as 
 \begin{equation}
\rm C=f^{(n_{T}-1)}=\frac{k_{1}^{(n_{T}-1)}}{k_{1}^{(0)}}.
\end{equation}
Here we mention that for the negative cooperative binding, 
$\rm f^{(n_{T}-1)}$ can be much less than $1$ in general. 
But here we consider the case $\rm X^{(0)}f^{(n_T-1)} >> 1.$

\end{document}